\documentstyle[PASJadd,12pt,psfig]{PASJ95}

\begin{document}

\title{Discovery of a Slow X-Ray Pulsator, AX~J1740.1$-$2847, in the Galactic Center Region}

\author{Masaaki {\sc Sakano} and Ken'ichi {\sc Torii}\\
{\it Space Utilization Research Program (SURP),
 National Space Development Agency of Japan (NASDA),}\\
{\it
 2-1-1 Sengen, Tsukuba, Ibaraki 305-8505}\\
{\it E-mail(MS): sakano@cr.scphys.kyoto-u.ac.jp}
\\[6pt]
Katsuji {\sc Koyama}\thanks{%
CREST: Japan Science and Technology Corporation (JST)}\\
{\it Department of Physics,
 Kyoto University,
 Kitashirakawa-Oiwake-cho, Sakyo, Kyoto 606-8502}
\\[6pt]
Yoshitomo {\sc Maeda}\\
{\it Department of Astronomy and Astrophysics,
 Pennsylvania State University,}\\
{\it
 University Park, PA 16802-6305, U.S.A.}\\
and \\
Shigeo {\sc Yamauchi}\\
{\it Faculty of Humanities and
 Social Sciences, Iwate University, 3-18-34 Ueda, Morioka, Iwate
 020-8550}
}

\abst{
   We report the discovery of an X-ray pulsar AX~J1740.1$-$2847 from the Galactic center
 region.   This source was found as a faint hard X-ray object on 7--8
 September 1998 with the {\it ASCA} Galactic center survey observation.   Then,
 coherent pulsations of $P=729\pm 14$~sec period were detected.  The
 X-ray spectrum is described by a flat power-law of $\simeq 0.7$ photon
 index.  The large absorption column of $\log N_{\rm H}\simeq$ 22.4
 (cm$^{-2}$) indicates that AX~J1740.1$-$2847 is a distant source, larger than 2.4
 kpc, and possibly near at the Galactic center region. The luminosity in the 2--10
 keV band is larger than 2.5$\times 10^{33}$ ${\rm erg}~{\rm s}^{-1}$, or likely to be $3.2\times 10^{34}$ ${\rm erg}~{\rm s}^{-1}$
 at the Galactic center distance.    Although the slow pulse period does not
 discriminate whether AX~J1740.1$-$2847 is a white dwarf or neutron star binary,
 the flat power-law and moderate luminosity
 strongly favor a neutron star binary.
}

\kword{\protect Pulsars --- Pulsars: individual (AX~J1740.1$-$2847)
--- Star: X-rays --- X-rays: binaries}

\maketitle
\thispagestyle{headings}

\section{Introduction}

     X-ray binary pulsars (here XBPs) constitute a bright class of X-ray
 sources in the sky, and extensive studies on XBPs have been so far
 made.  Still our knowledge on the XBP populations is far from complete,
 in particular, on transient and low luminosity XBPs in our Galactic
 plane and Galactic center regions, due either to limited observation time, poor
 spatial resolution of hard X-ray instruments, or low-energy limited
 band of X-ray telescopes onboard the previous satellites.
 As for transient XBPs,  Koyama et al.\ (1990) reported four XBPs and three
 possible candidates from the Scutum arm.
  Recent X-ray satellites,
 {\it RXTE} and {\it Beppo-SAX} have further found many transient XBPs
 from the Galactic Plane (e.g., Bildsten et al.\ 1997; Nagase 2000).  This, together with the {\it ASCA}
 discoveries of fainter  XBPs on the Galactic plane (e.g., Torii et al.\ 1998;
 Nagase 2000), suggests that a large fraction of faint and/or transient
 XBPs are still hidden in the Galactic plane and the center regions.

     This paper reports on the discovery of a very faint and slowly rotating
 X-ray pulsator AX~J1740.1$-$2847\footnote{This source was first designated as
 AX~J1740.2$-$2848 by Sakano \& Koyama (2000).}.  Based on the timing
 and spectral analyses, the nature and origin are discussed.

\section {Observations}

   An {\it ASCA} pointing observation with the field center at ($l_{\rm
 II}$, $b_{\rm II}$)=(359$^{\circ}\!$.49, 0$^{\circ}\!$.98) was made on
 7--8 September 1998 as a part of the {\it ASCA} Galactic plane / Galactic center
 survey project.  {\it ASCA} is equipped with two kinds of X-ray detectors,
 two Gas Imaging Spectrometers (GISs: Ohashi et al.\  1996; Makishima
 et al.\  1996), and two Solid-state Imaging Spectrometers (SISs: Burke
 et al.\  1991, 1994; Yamashita et al.\  1997), at the foci of four
 identical X-Ray Telescopes (XRTs: Serlemitsos et al.\  1995).

   The two GISs (GIS2 and GIS3) observed the same sky region
 simultaneously with $\simeq 50'$ diameter circular field of view.  The
 GIS was operated in PH nominal mode, which utilizes time resolutions of
 1/16 and 1/2~sec for high and medium bit rate data, respectively.

   The two SISs (SIS0 and SIS1), both comprised four CCD chips (C0--C3),
 were operated in complementary two-CCD mode with a time resolution of
 8~sec; SIS0 observed a rectangular sky of $11'\times 22'$ by using the
 two chips C1 and C2, while SIS1 observed the adjacent sky region of
 $11'\times 22'$.  Thus, together with the two SISs, a square region of
 $22'\times 22'$ is covered with the field center at the same position
 of the GIS field center.  As a result, the relevant source AX~J1740.1$-$2847 (see
 next section) was not located in the SIS0 field.  We have made the
 standard screening for all the data, then obtained the effective
 exposure times of 10.0 and 8.9 ksec for the GIS and SIS, respectively.

\section{Analysis and Results}

     We have made the GIS contour images (GIS2+GIS3) in the two energy bands of
 0.7--3 keV (the soft band) and 3--10 keV (the hard band) in figures 1a
 and b.
  We detected 3 point sources above the 5-sigma
 detection criterion; two of them are hard sources, while the other is
 a soft source.
     The positions of the two hard sources were determined to be
 ($17^{\rm h}$ $40^{\rm m}$ 11$^{\rm s}\hspace{-5pt}.\hspace{2pt}$6,
 $-28^\circ$ $47'$ $48''$)(J2000)  and 
 ($17^{\rm h}$ $40^{\rm m}$ 17$^{\rm s}\hspace{-5pt}.\hspace{2pt}$7,
  $-29^\circ$ $3'$ $56''$) (J2000) with respective 90\% error radii 
 of 30 and 45 arc seconds.
  No cataloged X-ray source is found within the error circles, hence 
 are designated to be AX~J1740.1$-$2847 and AX J1740.2$-$2903.
   The position of the other (soft source) is
 ($17^{\rm h}$ $40^{\rm m}$ 25$^{\rm s}\hspace{-5pt}.\hspace{2pt}$2,
  $-28^\circ$ $56'$ $52''$) (J2000) with the 90\% error radius of 45 arc seconds.
 Because of the positional coincidence and the apparent softness,
 this source is probably a ROSAT source, 1RXS J174024.6$-$285706 
(Voges et al.\ 1999, 2000).

  To all the detected sources, we made FFT analysis
 after barycentric correction for the photon arrival times.  The data
 used were accumulated from a circle of $3'$ radius around each source
 position, and the two GIS data were added to increase statistics.  We
 then found coherent pulsations from AX~J1740.1$-$2847.
 Here and after, we concentrate on the analysis of AX~J1740.1$-$2847.


   Figure~2 shows the power spectrum of AX~J1740.1$-$2847
 for the 1.4--9.0 keV band.  The maximum power peak corresponds
 to a period of $P\simeq 728$~sec.  We folded the same GIS data around
 this trial value and determined more accurate period of $P=729\pm
 14$~sec.  The rather large error is due to the long pulse period compared
 to the short observation time span of about 19.1 ksec.


  We folded the light curve of AX~J1740.1$-$2847 with the 729~sec
 period, and the resultant pulse profile is shown in
 figure~3.  With the same period, we also folded background
 data taken from the nearby source free region, and confirmed to exhibit
 no modulation.  The background level thus estimated is indicated by the
 dotted histogram in figure~3.
  The mean count rate of the source with one GIS detector after background
 subtraction is 2.0$\times 10^{-2}$ c~s$^{-1}$.
  The pulse profile is
 characterized by a single (main) peak, with a weak (sub) peak at the
 middle of the pulse phase.  The amplitude of the main pulse peak is
 nearly 100\% of the source flux.  We further made FFT and folding
 analysis for the SIS1 data (note that the source was not in the SIS0
 field).  Although the SIS result confirmed that of the GIS, no tighter
 constraint on the pulse period was obtained due to much reduced SIS
 photons than those from GIS.

  Apart from the pulsation, the GIS light curve of AX~J1740.1$-$2847 is
 found to show significant aperiodic variabilities.
 The RMS variation of the light curve in the 1.4--9.0 keV energy band
 is $\sim$40\% for the timescale of 729~sec.


    As is shown in figure~1a, AX~J1740.1$-$2847 is essentially absent in the soft
 band image, indicating a highly absorbed spectrum.  Using the data from
 the same areas for the source and background as those of the timing
 analysis, and summing the GIS2 and GIS3, we made the GIS spectrum as is
 shown in figure~4.  To this spectrum, we fitted conventional models; a
 power-law, thermal bremsstrahlung, and thin thermal model, each with
 absorption column.  All these models were acceptable, but 
 a bremsstrahlung and a thin thermal plasma models gave the best-fit
 temperature to be unrealistically high, higher than 40~keV.  Thus, a
 power-law model is preferred.


    The best-fit power-law model parameters were: photon index of
 $\Gamma$=0.7$^{+0.6}_{-0.6}$, hydrogen column density of $N_{\rm
 H}$=2.5$^{+2.9}_{-1.8}\times 10^{22}$ ${\rm H}~{\rm cm}^{-2}$, and normalization at
 1~keV of 1.6$^{+3.7}_{-1.0}\times 10^{-4}$
 photons~s$^{-1}$~cm$^{-2}$~keV$^{-1}$.  Then, the 2--10 keV band flux,
 after removing the absorption effect, was 4.1$\times 10^{-12}$
 ${\rm erg}~{\rm s}^{-1}~{\rm cm}^{-2}$.  The errors quoted here and after are 90\% confidence
 level.  Figure~4 shows the GIS2+3 spectrum and the best-fit
 power-law model convolved with the detector response. 
     No apparent iron K-line was found within the limited statistics.
 The upper limit of the equivalent width for a narrow line at
 6.4--6.7~keV was estimated to be 500~eV.

    To restore the secular degradation of the SIS CCDs, the
 pixel-to-pixel fluctuation of residual dark currents was removed (the
 RDD correction, see Dotani et al.\ 1997), where the residual dark
 distribution (RDD) maps were obtained from the data of 2-CCD mode
 operated within 1~month of the source observation.  The photon number
 thus corrected was 513 in the $3'$-radius circular region in S1C2, of
 which 56~\% was estimated to be background.  Using these data, we made
 the SIS1 spectrum and fitted it with a power-law model. The best-fit
 photon index and absorption column are $\Gamma= 0.1\pm 0.7$ and $N_{\rm
 H} = (7\pm 21)\times 10^{21}$ ${\rm H}~{\rm cm}^{-2}$, respectively.  Therefore, the
 SIS1 spectrum does not give more precise constraint than, but is
 consistent with, the GIS results.

\section{Discussion}

    The coherent slow pulsation of 729~sec period is consistent with a
 spin rotation of both a white dwarf
 (e.g., Patterson 1994)
 and a neutron star
 (e.g., Bildsten et al.\ 1997).  Therefore,
 AX~J1740.1$-$2847 would be a binary of white dwarf or neutron star (XBP).
   White dwarf binaries exhibit spectra of optically thin thermal
   plasma, which are characterized by strong emission lines from highly
   ionized iron (e.g., Ezuka \& Ishida 1999).  
 Neutron star binaries, on the other hand, generally exhibit
power-law spectra with a break at high energies (e.g., Nagase 1989).
 The best-fit spectrum gives no strong constraint on the iron line.
  The flat power-law spectrum, however, favors the XBP scenario for AX~J1740.1$-$2847,
although the large errors of the best-fit parameters  do not completely exclude a
white dwarf possibility.

     If AX~J1740.1$-$2847 is a neutron star binary, the low X-ray flux means that
 mass accretion rate from a companion, hence circumstellar absorption
 column, is small.  Therefore, a large fraction of the absorption column
 of AX~J1740.1$-$2847 is attributable to interstellar matter.  Sakano (2000) made a
 simple interstellar mass distribution model in the Galactic plane,
 combining several published results on HI and H$_2$ distribution, and
 obtained overall fit to the absorption columns of X-ray sources located
 in the direction to the Galactic center region.  Using this model and the
 absorption column of 2.5$^{+2.9}_{-1.8} \times  10^{22}$ ${\rm H}~{\rm cm}^{-2}$, we
 estimate the distance of AX~J1740.1$-$2847 to be consistent with that of the Galactic center,
 or larger than 2.4 kpc.
   Then, the
 2--10 keV flux of 4.1$\times 10^{-12}$ ${\rm erg}~{\rm s}^{-1}~{\rm cm}^{-2}$ is converted to
 3.2$\times 10^{34} {d_{\rm 8.5kpc}}^2$ ${\rm erg}~{\rm s}^{-1}$ for the intrinsic luminosity,
 where $d_{\rm 8.5kpc}$ is the source distance in unit of 8.5~kpc.
 This luminosity is
 in the range of XBPs (10$^{33}$--10$^{38}$ ${\rm erg}~{\rm s}^{-1}$;
 e.g., Negueruela et al.\ 2000; Stella et al.\ 1986), but is brighter than that of most white dwarf binaries
 ($\ltsim$ a few $\times 10^{33}$ ${\rm erg}~{\rm s}^{-1}$; e.g., Ezuka \& Ishida 1999).

     Consequently, all the results, the spectral shape, equivalent width
 of iron line and luminosity, favor that AX~J1740.1$-$2847 is an XBP rather than a
 white dwarf binary.  The long spin period of AX~J1740.1$-$2847 resembles that of
 the prototype system of persistent Be/X-ray binaries, X~Persei ($P
 \simeq 840$~sec; White et al.\  1976).  The hard spectrum and the low X-ray luminosity
 also resemble to X~Persei.   The most probable spin period is obtained by
 the condition that the co-rotation radius is equal to the Alfven
 radius, where no angular momentum is transferred from the neutron star
 to the accreting gas or vice versa.  Since the Alfven radius increases
 as the accreting gas decreases, a longer spin period requires a lower
 luminosity (e.g., Stella, White, \& Rosner 1986).  This mechanism
 should be working in the low-luminosity and long-period XBPs, such as
 X~Persei, AX~J170006$-$4157 ($P=$714.5~sec; Torii et al.\  1999), and the new X-ray pulsar AX~J1740.1$-$2847.

\par
\vspace{1pc}\par
   The authors express their thanks to all the members of the {\it ASCA}
 team.
   We are grateful to Prof. Nagase and an anonymous referee for their
 valuable comments and suggestions.
  They thank the help of the {\it ASCA} Galactic plane/center
 survey team.

\section*{References}
\small

\re
Bildsten L., Chakrabarty D., Chiu J., Finger M.H., Koh D.T., Nelson R.W., Prince T.A., Rubin B.C.\ et al.\ 1997, ApJS 113, 367

\re
Burke B.E., Mountain R.W., Harrison D.C., Bautz M.W., Doty J.P., Ricker G.R., Daniels P.J.\ 1991, IEEE Trans.\ ED-38, 1069

\re
Burke B.E., Mountain R.W., Daniels P.J., Dolat V.S.\ 1994, IEEE Trans.\ Nuc.\ Sci.\ 41, 375

\re
 Dotani T., Yamashita A., Ezuka H., Takahashi K., Crew G., Mukai K.,  the SIS Team 1997,
The {\it ASCA} news No. 5, 14 (Greenbelt: NASA GSFC)

\re
 Ezuka H., Ishida M.\ 1999, ApJ 120, 277

\re
Koyama K., Kawada M., Kunieda H., Tawara Y., Takeuchi Y., Yamauchi S.\ 1990, Nature 343, 148

\re
 Makishima K., Tashiro M., Ebisawa K., Ezawa H., Fukazawa Y., Gunji S., Hirayama M., Idesawa E.\ et al.\ 1996, PASJ 48, 171

\re
 Nagase F.\ 1989, PASJ 41, 1

\re
 Nagase F.\ 2000, Astrophysical Letters and Communications, in press

\re
Negueruela I., Reig P., Finger M.H., Roche P.\ 2000, A\&A 356, 1003

\re
 Ohashi T., Ebisawa K., Fukazawa Y., Hiyoshi K., Horii M., Ikebe Y., Ikeda H., Inoue H.\ et al.\ 1996, PASJ 48, 157

\re
 van Paradijs J. 1995, in X-ray Binaries,
 ed W.H.G.\ Lewin, J.\ van Paradijs, E.P.J.\ van den Heuvel, 
 (Cambridge UP), 536

\re
 Patterson J.\ 1994, PASP 106, 209

\re
 Sakano M.\ 2000, Ph.D.\ thesis, Kyoto University

\re
 Sakano M., Koyama K.\ 2000, IAU Circ.\ 7364

\re
 Serlemitsos P.J., Jalota L., Soong Y., Kunieda H., Tawara Y., Tsusaka Y., Suzuki H., Sakima Y.\ et al.\ 1995, PASJ 47, 105

\re
 Stella L., White N.E., Rosner R.\ 1986, ApJ 308, 669

\re
 Torii K., Kinugasa K., Katayama K., Kohmura T., Tsunemi H., Sakano M., Nishiuchi M., Koyama K., Yamauchi S.\ 1998, ApJ 508, 854

\re
 Torii K., Sugizaki M., Kohmura T., Endo T., Nagase F.\ 1999, ApJ 523, L65

\re
 Voges W., Aschenbach B., Boller T., Br\"{a}uninger H., Briel U., Burkert W., Dennerl K., Englhauser J.\ et al.\ 1999, A\&A 349, 389

\re
 Voges W., Aschenbach B., Boller T., Br\"{a}uninger H., Briel U., Burkert W., Dennerl K., Englhauser J.\ et al.\ 2000, IAU Circ.\ 7432

\re
 White N.E., Mason K.O., Sanford P.W., Murdin P.\ 1976, MNRAS 176, 201

\re
 Yamashita A., Dotani T., Bautz M., Crew G., Ezuka H., Gendreau K., Kotani T., Mitsuda K.\ et al.\ 1997, IEEE Trans.\ Nucl.\ Sci.\ 44, 847

\label{last}

\newpage
\section*{Figure Captions}


\begin{figure}[htbp]
\centering
\centerline{\psfig{file=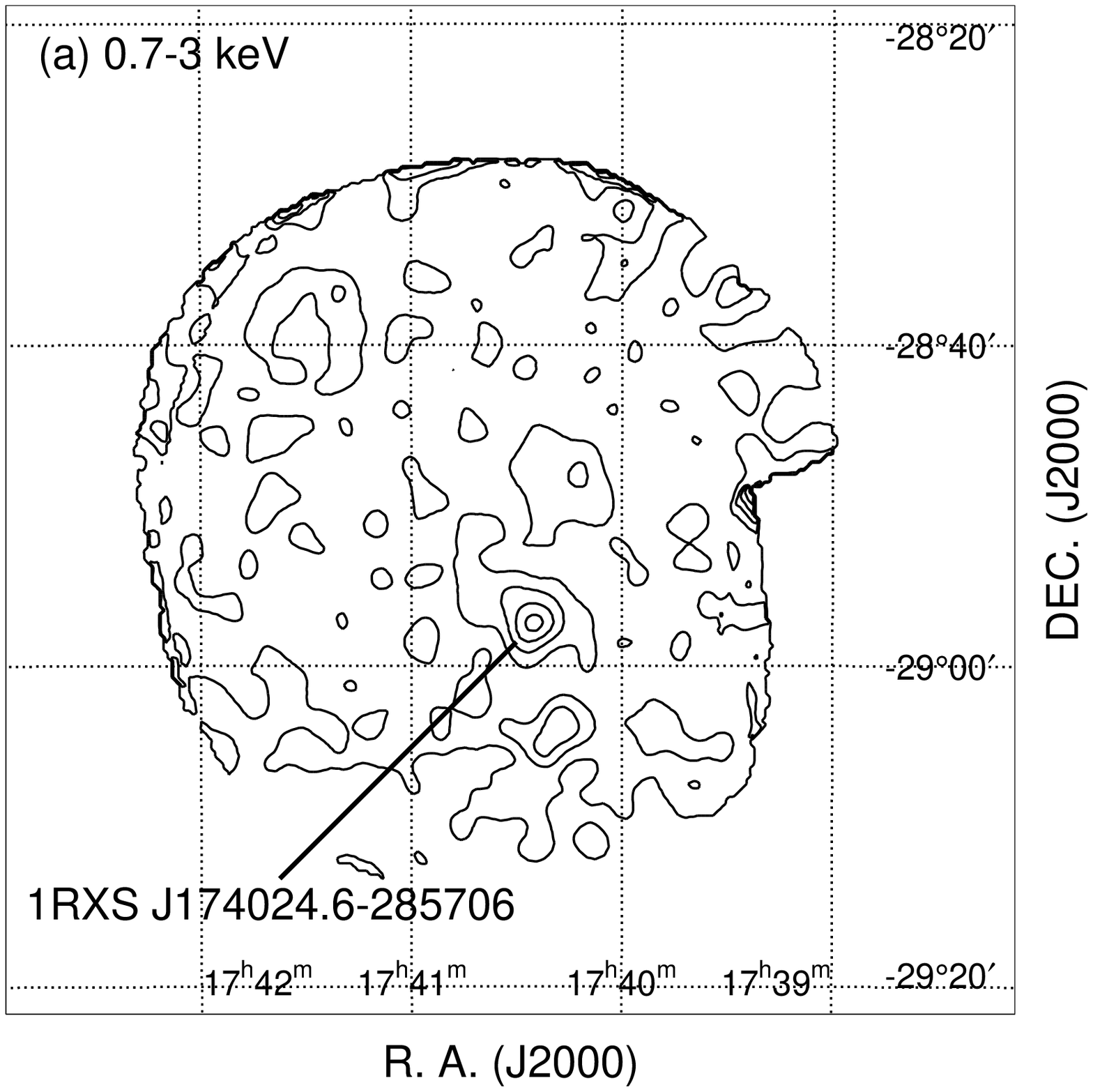,width=0.55\textwidth,clip=}}
\centerline{\psfig{file=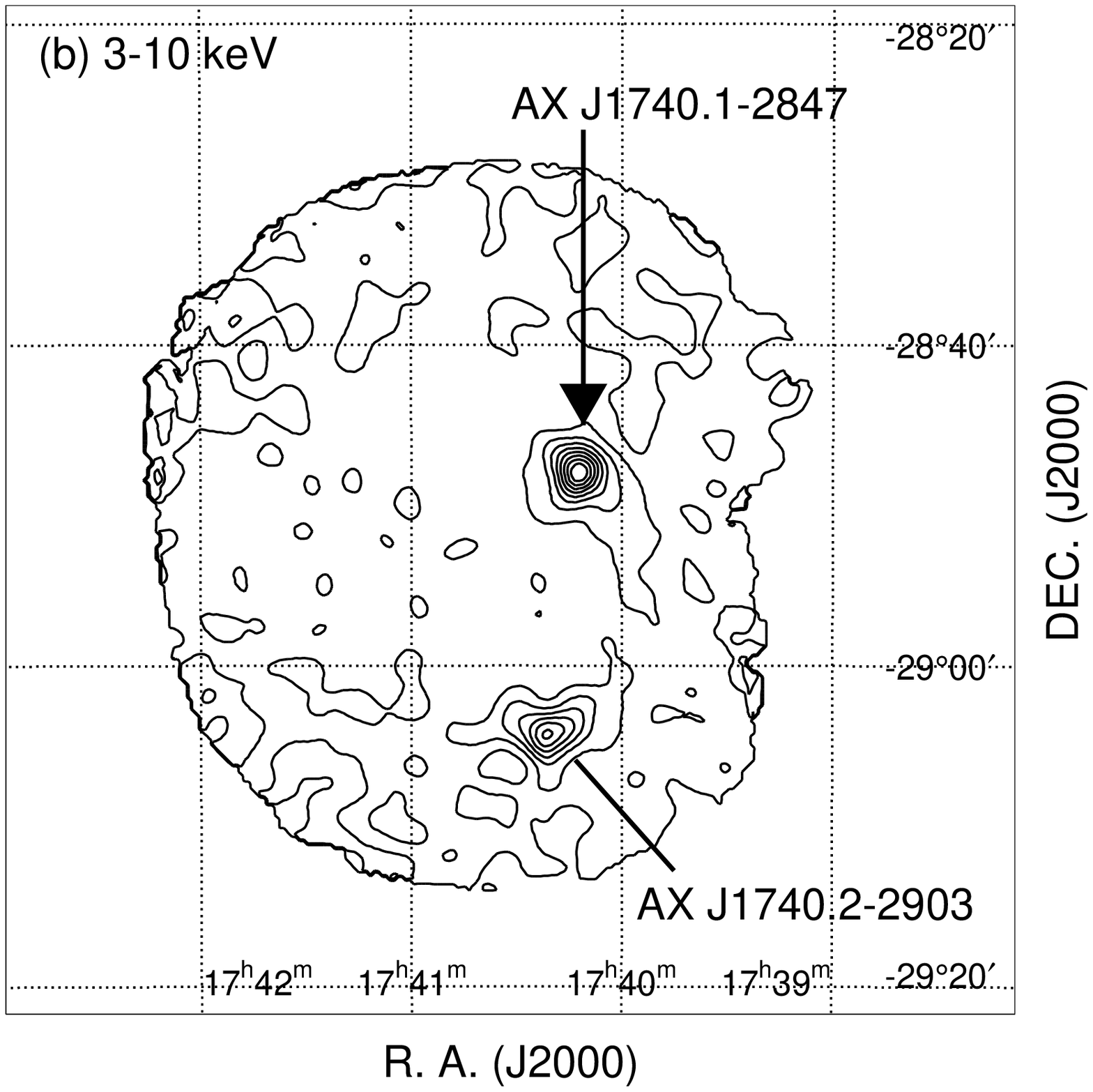,width=0.55\textwidth,clip=}}
\caption
{
 X-ray contour images for the (a) 0.7--3 keV and (b) 3--10 keV bands.
The data of GIS2 and GIS3 are summed,
 smoothed with a Gaussian filter of $\sigma =$ 3 pixels
 ($\sim$ 0.75 arcmin),
 and corrected for exposure, vignetting and the detection efficiency
 with GIS grid, after non X-ray background is subtracted.
 Contour levels are linearly spaced.}
\label{fig1}
\end{figure}


\begin{figure}[htbp]
\centering
\centerline{\psfig{file=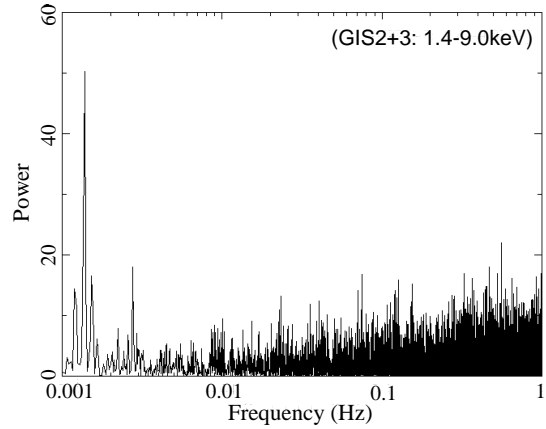,width=0.4\textwidth,clip=}}
\caption{
 The power spectrum of AX~J1740.1$-$2847 with the 1.4--9.0 keV band is shown. The power is normalized
so that the random fluctuation has an averaged value of 2.}
\label{fig2}
\end{figure}


\begin{figure}[htbp]
\centering
\centerline{\psfig{file=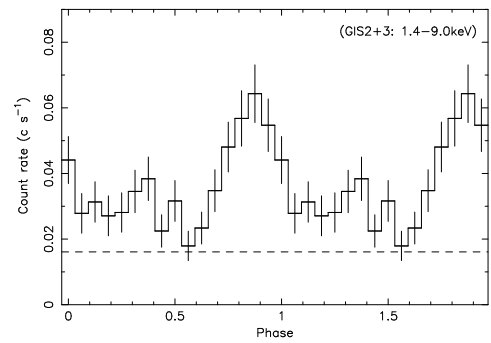,width=0.4\textwidth,clip=}}
\caption{
 The folded light curve of AX~J1740.1$-$2847 for the period of 729 s
 with the energy range of 1.4--9.0~keV is shown.
 The background level is given by a dashed line.}
\label{fig3}
\end{figure}


\begin{figure}[htbp]
\centering
\centerline{\psfig{file=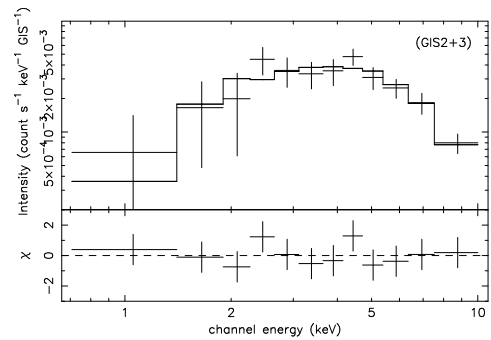,width=0.4\textwidth,clip=}}
\caption{
The X-ray spectrum of AX~J1740.1$-$2847.
 The
spectrum is fitted with a power-law function modified by interstellar
absorption. The residuals are shown in the bottom panel.}
\label{fig4}
\end{figure}

\end{document}